\documentclass[conference]{IEEEtran}
\ifCLASSINFOpdf
\else
\fi
\usepackage{balance}
\usepackage{pifont}

\usepackage{amsmath,amssymb,amsfonts}
\usepackage{graphicx}
\usepackage{textcomp}
\usepackage{xcolor}
\usepackage{xspace}
\usepackage{balance}
\usepackage{pifont}
\newcommand{\tool}{\textit{FEAD}\xspace}

\usepackage{algorithm}
\usepackage{algpseudocode} 
\usepackage{amsmath}
\usepackage{amssymb}
\usepackage{multirow}
 \usepackage{array} 
 \usepackage{enumitem}
 \usepackage{framed}
 \usepackage{makecell}
 \usepackage[misc]{ifsym}
 
 \usepackage{amssymb}
\usepackage[colorlinks=true, linkcolor=black, urlcolor=black, citecolor=black]{hyperref}

\begin{document}
%
\title{Attack Effect Model based Malicious Behavior Detection}


\author{
    Limin Wang, Lei Bu$^{(\textrm{\Letter})}$, Muzimiao Zhang, Shihong Cang, Kai Ye\\
	\IEEEauthorblockA{
     \textit{State Key Laboratory of Novel Software Techniques, Nanjing University, Nanjing, Jiangsu 210023, China}\\
     Email: bulei@nju.edu.cn, 
    }\vspace{-8.8mm}

}

\IEEEoverridecommandlockouts
\makeatletter\def\@IEEEpubidpullup{6.5\baselineskip}\makeatother
\IEEEpubid{\parbox{\columnwidth}{
		Network and Distributed System Security (NDSS) Symposium 2025\\
		24-28 February 2025, San Diego, CA, USA\\
		ISBN 979-8-9894372-8-3\\
		https://dx.doi.org/10.14722/ndss.2025.[23$|$24]xxxx\\
		www.ndss-symposium.org
}
\hspace{\columnsep}\makebox[\columnwidth]{}}

\maketitle

\begin{abstract}
Traditional security detection methods struggle to keep pace with the rapidly evolving landscape of cyber threats targeting critical infrastructure and sensitive data. These approaches suffer from three critical limitations: non-security-oriented system activity data collection that fails to capture crucial security events, growing security monitoring demands that lead to continuously expanding monitoring systems, thereby causing excessive resource consumption, and inadequate detection algorithms that result in the inability to accurately distinguish between malicious and benign activities, resulting in high false positive rates.

To address these challenges, we present FEAD
, an attack detection framework that improves detection by focusing on identifying and supplementing security-critical monitoring items and deploying them efficiently during data collection, as well as the locality of potential anomalous entities and their surrounding neighbors during anomaly analysis. \tool incorporates three key innovations:
(1) an attack model-driven approach that extracts security-critical monitoring items from online attack reports, enabling a more comprehensive monitoring items framework; (2) an efficient task decomposition mechanism that optimally distributes monitoring tasks across existing collectors, maximizing the utilization of available monitoring resources while minimizing additional monitoring overhead; (3) a locality-aware anomaly analysis technique that exploits the characteristic of malicious activities forming dense clusters in provenance graphs during active attack phases, guiding a vertex-level weight mechanism in our detection algorithm to better distinguish between anomalous and benign vertices, thereby improving detection accuracy and reducing false positives.

Evaluations show \tool outperforms existing solutions with an 8.23\% higher F1-score and 5.4\% overhead. Our ablation study also confirms that \tool’s focus-based designs significantly boost detection performance.


\end{abstract}

\IEEEpeerreviewmaketitle

\section{Introduction}

Modern computing systems across all scales are increasingly targeted by advanced cyber threats such as APT attacks~\cite{log4j}, posing significant security challenges. Traditional security measures often fail to address these evolving threats, prompting academia and industry to focus on lightweight attack detection systems. These systems typically utilize runtime log auditing and analysis (e.g., syscall logs)\cite{zhu2021general,zhu2023aptshield} to achieve effective security monitoring and attack detection. Among these, anomaly-based detection approaches\cite{yang2022systematic} have emerged as a promising solution by analyzing historical logs to identify deviations from established benign patterns~\cite{zhu2023aptshield,yang2023prographer,ding2023airtag}, thus effectively addressing the challenge of detecting novel and evolving attack variants.
While these approaches show promise, they face three critical challenges:

\noindent\ding{182} \textbf{Lack of Security-Oriented Efficient Monitoring.}
Existing monitoring tools (e.g., auditd~\cite{auditd,michael2020forensic,kurniawan2022krystal}) are primarily designed for performance or fault diagnosis rather than security. They focus on system failures and resource usage, offering limited visibility into security-relevant activities. Although efforts like eAudit~\cite{sekar2024eaudit} try to improve auditd by using eBPF and incorporating capabilities from existing monitors (e.g., Trace~\cite{irshad2021trace}), they still lack the capability to capture higher-level security events. Enterprise tools such as auditbeat~\cite{auditbeat} provide some enhancements but heavily rely on expert tuning. The absence of a systematic framework for identifying critical security monitoring points leads to incomplete coverage and security blind spots.


\noindent\ding{183} \textbf{High Deployment Costs of Security Monitoring Tasks.}
Evolving threats demand frequent updates to monitoring strategies, often requiring new monitoring modules for each new security monitoring requirement and increasing system complexity. Moreover, no single tool covers all needs, forcing organizations to deploy multiple systems, resulting in redundancy and performance degradation. 

\noindent\ding{184} \textbf{High False Positive Rates in Anomaly-Based Attack Detection.}
Current detection systems often struggle to accurately differentiate between legitimate system changes and genuine security threats, as benign and malicious nodes frequently exhibit similar behavioral patterns in monitoring data. For instance, routine updates or maintenance activities that alter network traffic may closely resemble attack signatures, resulting in false alarms. Monitoring blind spots further exacerbate this issue by limiting visibility and degrading the system's ability to distinguish normal variations from actual threats. This poor discrimination leads to alert fatigue among security teams and reduces the practical utility of security monitoring in enterprise environments.

To address these challenges, we present \tool (Focus-Enhanced Attack Detection), a framework that improves detection by focusing on two key aspects: identifying and supplementing security-critical monitoring items, and then deploying them efficiently during data collection, as well as analyzing the locality of potential anomalous entities and their surrounding neighbors during anomaly detection. This dual focus ensures targeted, effective attack detection while minimizing system overhead. Specifically, 
\textbf{For challenge 1}, \tool uses an \emph{Attack Effect Model} and large language models to analyze attack reports, breaking down attack steps and assessing their impact on systems and software to identify key monitoring items. This approach ensures comprehensive monitoring by systematically extracting monitoring requirements from real-world attacks.
\textbf{For challenge 2}, \tool introduces a novel task decomposition mechanism that breaks down complex tasks and distributes them across existing collectors, thereby collaboratively achieving target monitoring objectives. This maximizes the use of built-in monitoring capabilities while minimizing new module deployment, reducing system overhead.
\textbf{For challenge 3}, \tool exploits attack locality in provenance graphs, where malicious activities form dense clusters during active phases and sparse ones across other phases. This insight guides a vertex-level weighting mechanism in the detection algorithm, focusing on anomalous vertices and their neighbors. Unlike previous methods that analyze all vertices indiscriminately, \tool’s approach enhances discrimination between benign and malicious behaviors, targets high-risk areas, and improves accuracy while reducing false positives.

Evaluation demonstrates \tool’s superior performance, achieving an average 8.23\% higher F1-score compared to existing solutions, with a low overhead of 5.4\%. Our ablation study further validates the effectiveness of our focus-based design, showing that the integration of a security-oriented monitoring framework improves the F1-score by over 12.63\%, while leveraging attack locality patterns achieves a 9.52\% improvement. These results highlight \tool’s ability to enhance detection accuracy while maintaining system efficiency, making it a promising solution for real-world attack detection.

We summarize our core contributions as follows:
\begin{itemize}[leftmargin=*]
\item Propose an attack detection framework that integrates security-oriented monitoring and locality-aware anomaly analysis for enhanced detection accuracy.
\item Develop an efficient task decomposition mechanism to optimize monitoring coverage while minimizing overhead, achieving comprehensive security monitoring with 5.4\% overhead.
\item Evaluate \tool showing a 8.23\% improvement in F1-score over existing solutions, with ablation studies confirming significant gains from monitoring enhancements (12.63\%) and locality-based detection (9.52\%).
\end{itemize}

\section{Background and Related Work}
\subsection{Preliminaries}
\label{sec:pre}

\noindent\textbf{Attack Effect Model.} 
An \emph{Attack Effect Model} is defined as an ordered sequence of attack steps $T = [(T_1, E_1), (T_2, E_2), ..., (T_n, E_n)]$, where each $T_i$ represents a triple $\langle actor_i, action_i, target_i\rangle$ describing who performed what action on which target, and $E_i$ represents the corresponding system impact (e.g., Program Execution, File Modification). Each attack step $(T_i, E_i)$ can be mapped to a set of monitoring items $M_i$ required for detecting the system impact. The steps are ordered chronologically to reflect the temporal progression of the attack.


\noindent\textbf{Provenance Graph.} In system security monitoring, provenance graphs model system behaviors by transforming audit logs (e.g., Linux Auditd~\cite{auditd}) into a graph capturing causal relationships between system entities and activities. A provenance graph is a directed graph $G = (V, E)$, where $V$ represents system entities (e.g., processes, files, sockets) and $E$ represents interactions between them. Specifically, $V = \{v | v \in (Process \cup File \cup Socket)\}$, and $E \subseteq V \times V \times T$ captures relationships between entities, where $T$ denotes the type of entity behavior (e.g., read, write, execute). 
This structure enables detailed tracking of system behaviors, aiding in attack detection, forensic analysis, and attack investigation. 

\subsection{Threat Model}
Similar to prior research on provenance tracking and threat detection~\cite{alsaheel2021atlas,yu2021alchemist,dong2023we,shen2019attack2vec,zengy2022shadewatcher,wang2022threatrace,han2020unicorn}, we consider the OS kernel and security monitoring components as part of our trusted computing base (TCB). We assume that the collected provenance data is reliable and has not been tampered with by attackers. Although attackers may attempt to subvert the system or compromise the monitoring components, such subversion activities can be captured in logs before they are compromised. Our focus is on detecting attacks that exploit application vulnerabilities or leverage social engineering techniques to gain unauthorized access to victim systems for data exfiltration or manipulation, rather than hardware-based or side-channel attacks.

We assume that the system is initially in a benign state, with the attack originating from outside the enterprise network. Attackers typically gain initial access through remote network exploitation, compromised credentials, or social engineering, and proceed with multi-stage operations that may include information gathering, exploitation of vulnerable software, payload deployment, privilege escalation, and lateral movement. Our approach focuses on identifying these attack patterns within system behavior data.



\subsection{System Monitoring Tools}

We motivate our research with an study of existing system monitoring tools to identify their capabilities and limitations for security monitoring applications. 

\subsubsection{Systematic Tool Collection Using Snowball Technique}




System monitoring tools are continuously evolving and scattered across platforms, often lacking unified indexing. To address this, we developed a snowball-based retrieval methodology~\cite{nainna2024cyber,rahman2020literature}, consisting of two main phases.

This method began with \emph{Linux Auditd}~\cite{auditd} as the initial seed due to its prevalence as Linux's default audit tool. We constructed a structured query template combining \ding{172} scenario-related terms (e.g., "Provenance graph", "Causal graph"), \ding{173} functionality-related terms (e.g., "Data collection", "Log collection"), and \ding{174} tool names (e.g., "Auditd" ). We then executed systematic searches across multiple academic databases (i.e., \emph{Google Scholar}, \emph{IEEE Xplore}, and \emph{ACM Digital Library}) using this template (e.g., \emph{(Provenance graph OR causal graph OR forensic analysis OR investigation) AND (data collection OR log collection) AND Auditd}), and removed irrelevant literature to build a normalized database.

Two co-authors with 3–5 years of relevant experience then extracted monitoring tools from the collected literature. Newly discovered tools were added to the query template (i.e., component \ding{174}) and used in subsequent search iterations, thus expanding our search scope to capture literature referencing these newly discovered tools. The process terminated when no new tools appeared in two consecutive rounds or when the search queue was exhausted. This iterative approach effectively addressed the fragmentation and rapid evolution of monitoring tools.

\begin{table}[tb!]
\small
\setlength{\tabcolsep}{2pt} 
\centering
\caption{System Monitoring Tools Usage Statistics}
\scalebox{0.8}{
\begin{tabular}{|c|l|c|c|c|}
\hline
Monitoring Tool & Usage Instances & Count & Percentage & Platform \\
\hline
Auditd~\cite{auditd} & 
\makecell[l]{\cite{michael2020forensic,kurniawan2022krystal,hassan2020omegalog,yu2021alchemist,inam2022forensic,irshad2021trace,milajerdi2019poirot} \\
\cite{hossain2020combating,chen2021clarion,wang2020you,zengy2022shadewatcher,milajerdi2019holmes,hassan2020tactical}} & 13 & 28.26\% & Linux \\
\hline
CamFlow~\cite{pasquier2018ccs} & 
\makecell[l]{\cite{han2020unicorn,pasquier2019here,goyal2024r,ding2023airtag,cheng2024kairos,satapathy2023disprotrack} \\
\cite{wang2022threatrace,chen2021clarion,yu2024cost,jia2024magic,xie2020pagoda}}
& 11 & 23.91\% & Linux \\
\hline
ETW~\cite{etw} & 
\makecell[l]{\cite{kurniawan2022krystal,hassan2020tactical,milajerdi2019poirot,yang2020ratscope,wang2020you,milajerdi2019holmes} \\
\cite{yagemann2021validating,zeng2022palantir,cheng2024kairos}}
& 9 & 17.39\% & Windows \\
\hline
SPADE~\cite{gehani2012spade} & \cite{chen2021clarion,yu2021alchemist,inam2022forensic,wu2022paradise,hussain2023towards,yu2024cost} & 6 & 13.04\% & Linux \\
\hline
Auditbeat~\cite{auditbeat} & \cite{zhu2023aptshield,zeng2021watson} & 2 & 4.35\% & Linux \\
\hline
PASSv2~\cite{muniswamy2009layering} & \cite{xie2019p,xie2018pagoda} & 2 & 4.35\% & Linux \\
\hline
UBSI~\cite{irshad2021trace} & \cite{irshad2021trace} & 1 & 2.17\% & Linux \\
\hline
eAudit~\cite{sekar2024eaudit} & \cite{sekar2024eaudit}& 1 & 2.17\% & Linux \\
\hline
sysdig~\cite{sysdig} & \cite{fang2022back} & 1 & 2.17\% & Linux \\
\hline
strace~\cite{strace} & \cite{datta2022alastor} & 1 & 2.17\% & Linux \\
\hline
Total & - & 47 & 100\% & - \\
\hline
\end{tabular}
}
\label{tab:monitor}
\end{table}

\subsubsection{Tool Collection Results}

Our snowball search technique yielded comprehensive results as shown in Table~\ref{tab:monitor}. The statistics indicate that in the Linux ecosystem, \textit{Auditd} is the most widely used monitoring tool, appearing in 13 papers (28.26\%) and ranking first. \textit{CamFlow} follows closely, used in 11 papers (23.91\%). For Windows platforms, \textit{ETW} (Event Tracing for Windows) is the predominant tool, used in 9 papers (17.39\%).
From our analysis, we classified these monitoring tools into three main categories:
\begin{enumerate}
    \item \textbf{Whole-system Provenance Collection Tools} (\textit{CamFlow}, \textit{SPADE}, \textit{PASSv2}): These focus on system-level data flow and causality tracking. For example, \textit{CamFlow} implements efficient monitoring by integrating Linux Security Modules (LSM) and NetFilter, while \textit{SPADE} utilizes \textit{Linux Auditd} logs to build provenance graphs supporting distributed environments. \textit{PASSv2} is a layered provenance architecture based on Linux 2.6 kernel (circa 2009) that integrates provenance across multiple abstraction layers through a unified Disclosed Provenance API, demonstrating early approaches to cross-layer provenance collection despite being constrained by legacy technology.

    \item \textbf{Audit Tools} (\textit{Auditd}, \textit{Sysdig}, \textit{ETW}, \textit{Auditbeat}, \textit{eAudit}): These record system behaviors to support security analysis and compliance auditing. \textit{Auditd} is Linux's default audit framework, \textit{Sysdig} uses kernel modules for event capture, \textit{ETW} is Windows' standard audit tool, \textit{Auditbeat} extends \textit{Auditd} with modern features, and \textit{eAudit} combines \textit{Auditd} with \textit{eBPF} technology.

    \item \textbf{Fine-grained Information Collection Tools} (\textit{UBSI}, \textit{strace}): These focus on precise monitoring of specific system behaviors. \textit{UBSI} provides unit-level behavior monitoring through static analysis, while \textit{strace} records detailed system call-level interactions between processes and the operating system.
\end{enumerate}

In practical applications, whole-system provenance collection and audit tools can directly build provenance graphs, while fine-grained information collection tools typically supplement the former by providing detailed parameter information for key nodes in the provenance graph.

\subsubsection{Security Monitoring Capability Analysis}

\begin{table}[tb!]
\renewcommand{\arraystretch}{0.9}
\setlength{\tabcolsep}{2pt} 
\small
\centering
\caption{System Call and Relationship Scenario Classification}
\scalebox{0.77}{
\begin{tabular}{|c|l|p{3.3cm}|p{4cm}|l|}
\hline
\textbf{Scenario} & \textbf{No.} & \textbf{Monitoring Event} & \textbf{Description} & \textbf{Source} \\
\hline
\multirow{6}{*}{\parbox{1.3cm}{File\\Operations}} & 1 & read & Read file content & Auditd \\
& 2 & RL\_READ & Read inode & CamFlow \\
& 3 & write & Write file content & Auditd \\
& 4 & RL\_WRITE & Write inode & CamFlow \\
& 5 & open & Open file & Auditd \\
& 6 & close & Close file & Auditd \\
\hline
\multirow{8}{*}{\parbox{1.3cm}{Directory\\Operations}} & 7 & creat & Create new empty file & Auditd \\
& 8 & unlink & Delete file & Auditd \\
& 9 & link & Create hard link & Auditd \\
& 10 & linkat & Create relative path hard link & Auditd \\
& 11 & unlinkat & Delete file in relative directory & Auditd \\
& 12 & rmdir & Remove directory & Auditd \\
& 13 & mkdir & Create directory & Auditd \\
& 14 & RL\_INODE\_CREATE & Create inode & CamFlow \\
\hline
\multirow{7}{*}{\parbox{1.3cm}{Process\\Operations}} & 15 & fork & Create new process & Auditd \\
& 16 & clone & Create new process (shared address space) & Auditd \\
& 17 & execute & Execute new program & Auditd \\
& 18 & RL\_CLONE\_MEM & Memory copy during cloning & CamFlow \\
& 19 & RL\_SETUID & Set user ID & CamFlow \\
& 20 & RL\_SETGID & Set process group ID & CamFlow \\
& 21 & kill & Send signal & Auditd \\
\hline
\multirow{4}{*}{\parbox{1.3cm}{IO Control}} & 22 & RL\_READ\_IOCTL & IO control read operation & CamFlow \\
& 23 & RL\_WRITE\_IOCTL & IO control write operation & CamFlow \\
& 24 & pipe & Create pipe & Auditd \\
& 25 & fcntl & File control operation & Auditd \\
\hline
\multirow{15}{*}{\parbox{1.3cm}{Network\\Operations}} & 26 & socket & Create socket & Auditd \\
& 27 & RL\_SOCKET\_CREATE & Create socket & CamFlow \\
& 28 & \parbox{2.3cm}{RL\_SOCKET\_PAIR \\ \_CREATE} & Create socket pair & CamFlow \\
& 29 & connect & Connect to remote host & Auditd \\
& 30 & RL\_CONNECT & Socket connection operation & CamFlow \\
& 31 & RL\_BIND & Socket binding operation & CamFlow \\
& 32 & RL\_LISTEN & Socket listening operation & CamFlow \\
& 33 & RL\_ACCEPT & Socket accept connection operation & CamFlow \\
& 34 & sendto & Send data to specified address & Auditd \\
& 35 & recvfrom & Receive data from specified address & Auditd \\
& 36 & sendmsg & Send message & Auditd \\
& 37 & sendmmsg & Send multiple messages & Auditd \\
& 38 & recvmsg & Receive message & Auditd \\
& 39 & recvmmsg & Receive multiple messages & Auditd \\
& 40 & getpeername & Get remote address of connected socket & Auditd \\
\hline
\multirow{6}{*}{\parbox{1.3cm}{Memory\\Operations}} & 41 & dup & Duplicate file descriptor & Auditd \\
& 42 & dup2 & Duplicate file descriptor to specified descriptor & Auditd \\
& 43 & RL\_MMAP & Memory mapping mount & CamFlow \\
& 44 & RL\_MMAP\_PRIVATE & Private memory mapping mount & CamFlow \\
& 45 & RL\_SH\_READ & Shared memory read operation & CamFlow \\
& 46 & RL\_PROC\_READ & Read process memory & CamFlow \\
\hline
\multirow{2}{*}{\parbox{1.3cm}{Message\\Queue}} & 47 & mq\_open & Open message queue & Auditd \\
& 48 & RL\_MSG\_CREATE & Create message & CamFlow \\
\hline
\multirow{4}{*}{\parbox{1.3cm}{System\\Loading}} & 49 & RL\_LOAD\_FILE & Load file to kernel & CamFlow \\
& 50 & RL\_LOAD\_FIRMWARE & Load firmware to kernel & CamFlow \\
& 51 & RL\_LOAD\_MODULE & Load module to kernel & CamFlow \\
& 52 & RL\_VERSION & Connect entity object version & CamFlow \\
\hline
\end{tabular}
}
\label{tab:provmonitor}
\end{table}


To assess the monitoring capabilities of these tools, we selected two of the most widely used system monitoring tools and analyzed their monitoring capabilities.
Based on the statistical data in Table~\ref{tab:monitor}, \textit{Auditd} and \textit{CamFlow} are the most widely adopted tools for provenance graph construction. These two tools have complementary monitoring capabilities: \textit{Auditd} uses system call tracing mechanisms to monitor basic events such as file operations, process behaviors, and network communications, while \textit{CamFlow} leverages the LSM framework to provide more granular tracking of entity relationships, covering memory operations, IO control, and system loading scenarios.
We first categorized their monitoring capabilities into eight major dimensions: file operations, directory operations, process control, IO management, network communications, memory operations, message queues, and system loading.
Table~\ref{tab:provmonitor} shows a detail of the monitoring events covered by these tools. Then, through our analysis, we identified the monitoring focus and limitations of each tool.
Based on the comprehensive analysis of the monitoring events presented in Table~\ref{tab:provmonitor}, we can draw the following conclusions:

\smallskip
\noindent\textbf{Requirement Gaps: Original design objectives misaligned with security monitoring needs. } Our analysis reveals that existing tools demonstrate a significant misalignment with security monitoring requirements. As evident from Table~\ref{tab:provmonitor}, both \textit{Auditd} and \textit{CamFlow} were originally designed for general-purpose system monitoring rather than security-specific monitoring. The table shows that neither tool adequately covers critical security-relevant operations such as environment variable manipulations, which are frequently exploited in attacks~\cite{env}. 

\smallskip
\noindent\textbf{Fragmented Monitoring Ecosystem: No unified system provides comprehensive coverage.} As illustrated in Table~\ref{tab:provmonitor}, no single monitoring system comprehensively implements all necessary monitoring capabilities. For instance, \textit{Auditd} demonstrates weak coverage in the \emph{System Loading} dimension (items 49-52), while \textit{CamFlow} excels in this area. Conversely, \textit{CamFlow} shows incomplete monitoring for \emph{Directory Operations} (items 7-14), which \textit{Auditd} covers extensively. Our deeper investigation found that security-optimized monitoring tools like \textit{Auditbeat} have attempted to address these gaps by augmenting \textit{Auditd} with support for system events (user logins, etc.) and file integrity monitoring~\cite{auditbeat_ref}. However, these enhancements are primarily guided by expert experience rather than systematic methodology, resulting in inevitable blind spots for common security-relevant operations. For example, despite its security focus, \textit{Auditbeat} still cannot monitor environment variable manipulations. This fragmentation highlights the urgent need for a more systematic approach to security monitoring capability design.

Our analysis of existing monitoring tools reveals critical limitations that demand immediate attention. The requirement gaps and fragmented monitoring ecosystem pose significant challenges for effective security monitoring. If left unaddressed, these issues will lead to persistent monitoring blind spots, resulting in reduced accuracy during attack detection and analysis. Furthermore, as new monitoring requirements emerge, organizations are forced to either develop custom monitoring modules (increasing development costs) or deploy multiple overlapping systems simultaneously (causing redundant data collection and performance degradation). These challenges urgently necessitate:
\begin{itemize}
    \item 
\textbf{\emph{A unified monitoring framework}} that systematically covers security-relevant operations to eliminate blind spots and enhance the quality of collected data and resulting provenance graphs, thereby improving attack detection accuracy.

\item 
\textbf{\emph{A lightweight deployment solution}} that consolidates essential monitoring capabilities while minimizing resource consumption, enabling organizations to adapt to evolving security requirements without prohibitive operational overhead.
\end{itemize}

\subsection{Provenance Graph-based Anomaly Detection}
Provenance graphs are widely used for anomaly detection due to their ability to capture system behaviors and causal relationships. Early methods, such as StreamSpot~\cite{manzoor2016fast} and Unicorn~\cite{han2020unicorn}, relied on graph kernels for clustering, but struggled with stealthy threats and subtle structural differences in rare anomalies.
Machine learning-based methods~\cite{Liu2019Log2vecAH,du2017deeplog,zhang2019robust} aim to learn complex patterns from provenance graphs using only benign data, but often fail to capture critical structural information, leading to poor detection of subtle or novel threats.
With the rapid development of graph neural network (GNN) techniques~\cite{bessadok2022graph}, they have become a popular choice for anomaly detection in provenance graphs~\cite{wang2022threatrace,zengy2022shadewatcher,tang2023gadbench}. GNNs can capture complex structural patterns, adapt to dynamic systems, and scale to large datasets, improving detection accuracy. However, many GNN methods treat vertices and edges uniformly, making it difficult to distinguish between benign and malicious behaviors, leading to false positives. These challenges urgently necessitate:
\begin{itemize}
    \item \textbf{\emph{A context-aware detection framework}} that intelligently differentiates between normal and suspicious behaviors based on their semantic context and relationships.
\end{itemize}

\section{FEAD: Focus-Enhanced Attack Detection}

\begin{figure}[!tb]
\centering
\includegraphics[width=1\linewidth]{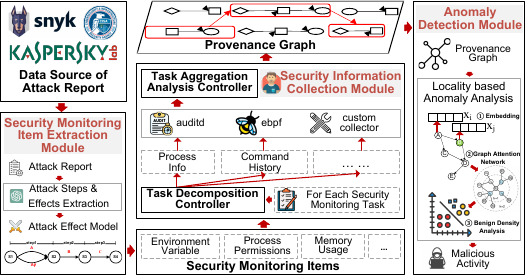}
\caption{The workflow of \tool.}
\label{fig:fead}
\end{figure}

As shown in Fig.~\ref{fig:fead}, \tool addresses the challenges of non-security-oriented data collection, high deployment costs, and high false positive rates in attack detection with three key components: \textbf{(1) The Security Monitoring Item Extraction Module}, which utilizes the \emph{Attack-Effect Model} to extract critical security monitoring items from attack reports by analyzing attack steps and impacts; \textbf{(2) The Security Information Collection Module}, which employs \emph{Adaptive Security Monitoring} to decompose complex tasks, distribute subtasks among existing collectors, and aggregate data for comprehensive analysis, minimizing system overhead; and \textbf{(3) The Anomaly Detection Module}, which applies \emph{Locality-based Anomaly Analysis} by leveraging the dense clustering of malicious activities while remaining sparse across different attack phases.

\subsection{The Security Monitoring Item Extraction Module}
\label{sec:monitor}
To tackle the lack of security-oriented efficient monitoring, we propose a systematic approach to identify security-critical monitoring items from online attack reports. This approach involves two main steps: \textbf{Attack Report Collection} and \textbf{Key Information Extraction}.

\subsubsection{Attack Report Collection}

Our attack report collection followed two main steps: \textbf{Report Crawling} and \textbf{Report Filtering}. 

\smallskip
\noindent\textbf{Report Crawling.} During the data acquisition phase, we crawled attack reports from multiple sources to build a comprehensive dataset.
Guided by previous studies \cite{sills2020cybersecurity,ladisa2023sok,li2022attackg}, we developed web crawlers to gather reports from platforms such as Snyk \cite{snyk}, Microsoft Security Intelligence Center \cite{microsoft}, and CISA \cite{cisa}, among others (due to space limitations, the complete list is available on our website \cite{fead}). To enhance coverage, we also included attack cases from the MITRE ATT\&CK knowledge base \cite{attack}.

\smallskip
\noindent\textbf{Report Filtering.} To ensure relevance and quality of collected reports, we rigorously filtered reports based on predefined criteria Table~\ref{tab:criteria}. This process involved collaboration among four authors (with 2–3 years of attack detection experience) and two industry experts (with 7–8 years in cybersecurity). After manual review, the final dataset comprised 260 APT reports and 7,098 attack cases spanning 268 MITRE ATT\&CK techniques (Ref.~\cite{fead} for detail).

\begin{table}[t]
\small
\centering

\caption{Inclusion and exclusion criteria}
\label{tab:report_criteria}
\small
\scalebox{0.9}{
\begin{tabular}{|p{0.065\textwidth}|p{0.4\textwidth}|}
\hline
\multicolumn{1}{|l|}{Type} & \multicolumn{1}{l|}{Description} \\ \hline
\noindent Inclusion & 
\begin{tabular}[t]{p{0.4\textwidth}}
    - Contains technical details on attack steps \\
    - Relevant to target systems, environments, or industry \\
    - Published within the last 5 years \\
\end{tabular} \\ \hline
\noindent Exclusion & 
\begin{tabular}[t]{p{0.4\textwidth}}
- Duplicate attack reports \\
- Reports with insufficient length (less than 200 words)\
\end{tabular} \\ \hline
\end{tabular}
}
\label{tab:criteria}
\end{table}


\subsubsection{Key Information Extraction}
Recent advancements in Large Language Models (LLMs) have made them highly effective for information extraction due to their vast knowledge and language understanding~\cite{li2023semi,sun2024umie,chen2022knowprompt}. However, LLMs still face challenges like hallucinations~\cite{ji2023survey,zhang2023siren}, which affect extraction accuracy. To address this, we use the Chain-of-Thought (CoT) prompting technique~\cite{feng2024towards,wei2022chain,chu2023survey}. By designing a reasoning process, we break down complex extraction tasks into manageable, smaller steps that LLMs can directly process, thereby reducing hallucinations and improving the transformation of unstructured attack reports into \emph{Attack Effect Model}
and reason out corresponding security monitoring items.

\begin{figure}[!tb]
\centering
\includegraphics[width=1\linewidth]{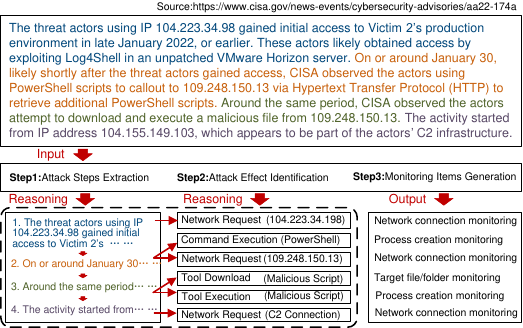}
\caption{The workflow of monitoring items generation.}
\label{fig:cot_process}
\end{figure}

In this work, we implement CoT by instructing LLM to follow our defined steps (in Fig.\ref{fig:cot_process}) while providing explicit reasoning for each step's output (Ref. \cite{fead} for detailed prompt).

\smallskip
\noindent\textbf{Step 1: Attack Steps Extraction}
We ask LLM to analyze the syntactic structure of input attack report text (I) to identify subject-verb-object relationships, forming triples (T = {$\langle$ $actor_i$, $action_i$, $target_i$$\rangle$}), where each triple represents a distinct attack step. 
As shown in the Fig.\ref{fig:cot_process}, from ``The threat actors using IP 104.223.34.98 gained initial access to Victim 2’s production environment", we form ($\langle$ \text{threat actors}, \text{Network Request}, \text{Victim 2’s production environment}$\rangle$).

\smallskip
\noindent\textbf{Step 2: Attack Effect Identification}
Attack effects represent the impact of an action on targets, we focus on ($\langle$ $action_i$, $target_i$$\rangle$) pairs from (T). For each pair, we further ask the LLM to analyze the context to determine its corresponding attack effect ($E_i$). For example, ($\langle$\text{Tool Execution}, \text{PowerShell}$\rangle$) triggers (E = \text{Program Execution}), indicating a behavioral effect on the system.

\smallskip
\noindent\textbf{Step 3: Monitoring Items Generation}
For each ($\langle$$action_i$, $target_i$, $E_i$$\rangle$), the LLM finally generates corresponding monitoring item ($M_i$). As shown, when (E = \text{Program Execution}), the LLM generates (M = \text{Process creation monitoring}).

By combining the attack report text (I) with our designed CoT prompt (Ref.~\cite{fead} for detail) as input to the LLM, we guide the LLM to decompose attack descriptions into structured steps, reason about their system impacts, and derive corresponding monitoring items.

\subsection{The Security Information Collection Module}
\label{sec:secmonitor}

In the previous section, we identified security-critical monitoring targets by analyzing attack reports and applying our attack effect model. However, identifying what to monitor is only part of the solution—implementing these monitors remains costly.

To address this, we propose a lightweight collaborative security monitoring architecture. It systematically decomposes complex monitoring tasks and intelligently assigns them to existing monitors. By integrating collected data to meet original goals, this approach reuses existing capabilities, reduces the need for new monitoring modules, and therefore minimizes performance overhead.

The following sections first introduce key \textbf{\textit{Definitions}}, then present our task decomposition, integration, and deployment methodology.

\subsection{Symbolic Definition of Security Monitoring Capabilities}
To accomplish our monitoring objectives, we establish a symbolic definition framework that standardizes the representation of existing monitoring tools' capabilities and our monitoring goals, enabling automated analysis, task decomposition, appropriate allocation of subtasks, and result integration.

Through analysis of existing information collection tools, we found that these tools typically collect data to build provenance graphs through specialized log parsing algorithms. Based on this observation, 
we adopt a symbolic representation format compatible with both these tools and provenance graphs, enabling a unified framework for defining monitoring capabilities.
This design ensures compatibility with existing systems while supporting standardized decomposition and integration of sub monitoring tasks.

Specifically, based on the \emph{Provenance Graph} model in Section~\ref{sec:pre}, we define the monitoring capability set as $C = \{c_1, c_2, ..., c_n\}$. Each monitoring capability $c_i$ is characterized by a triple $\langle V_c,O_c,T_c \rangle$:

\begin{itemize}
    \item $V_c$ represents the set of system entities observable by this monitoring capability
    \item $O_c$ represents property descriptions of monitored entities and their output category sets
    \item $T_c$ represents the set of system event types observable by this monitoring capability
\end{itemize}

For the entity property set $O_c$ in monitoring capability $c_i$, we define:
\begin{equation}
O_c = \{(a_1, t_1), (a_2, t_2), ..., (a_m, t_m)\}
\end{equation}
where $a_i$ represents the property of the monitored entity, and $t_i$ represents that property's data type. Specifically, property data types include:

\begin{itemize}
    \item Basic data types: integers ($\mathbb{Z}$), real numbers ($\mathbb{R}$), boolean values ($\{\text{true}, \text{false}\}$), strings ($\Sigma^*$)
    \item Composite data types: lists, sets, key-value pairs, etc.
    \item Time-series data types: representing continuously sampled metric values in form $(t, v)$, where $t$ represents a timestamp and $v$ represents the sampled value
\end{itemize}

To achieve our monitoring objectives, we designed logical operations to decompose complex tasks and integrate results. We define operators ($\lambda$) based on data types to flexibly combine existing system monitors. These logical operators ($\lambda$) are as follows:


\begin{itemize}
\item Logical operations: AND ($\land$), OR ($\lor$), NOT ($\neg$)
\item Set operations: Element contains ($\in$), Subset relationship ($\subseteq$), Union operation ($\cup$), Intersection operation ($\cap$)
\item String operations: String matching ($\text{match}$), String concatenation ($\text{concat}$), String splitting ($\text{split}$), Substring contains ($\text{contains}$)
\item Numeric operations: Greater than ($>$), Less than ($<$), Equal to ($=$), Sum operation ($\text{sum}$), Average operation ($\text{avg}$)
\end{itemize}

These operators ensures that after breaking tasks into subtasks for existing monitors to collect information and then integrating this data, the resulting monitoring entities and their properties align with our original monitoring goals.

\subsection{Lightweight Collaborative Security Monitoring Framework}

\begin{figure}[!tb]
\centering
\includegraphics[width=0.8\linewidth]{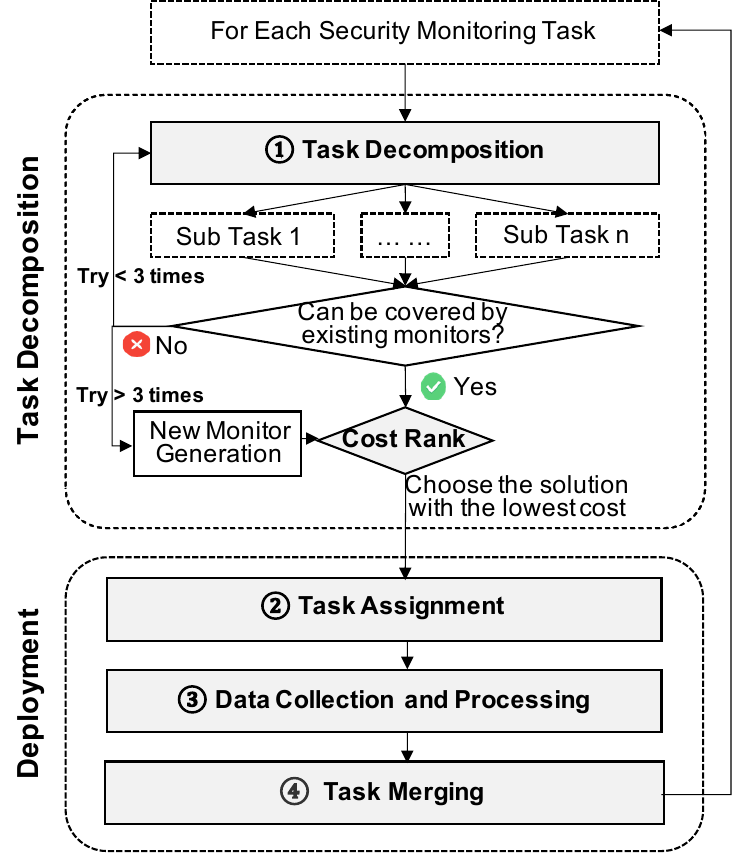}
\caption{The workflow of Lightweight Collaborative Security Monitoring Framework.}
\label{fig:light_weight_monitor}
\end{figure}

Based on the definitions established earlier, we propose a systematic task decomposition and data integration methodology as illustrated in Figure \ref{fig:light_weight_monitor}. Our lightweight collaborative security monitoring framework transforms complex monitoring requirements into manageable subtasks through recursive decomposition. The workflow begins with \textit{Task Decomposition} where security monitoring tasks are broken down and evaluated against existing capabilities. If existing monitors can solve the subtasks, a \textit{Cost Rank} function evaluates implementation solutions, selecting only the lowest-cost option. When existing capabilities cannot fulfill monitoring requirements after multiple attempts, \textit{New Monitor Recommendation} suggest appropriate new monitoring components. 
Once optimal solutions are identified, they proceed to \textit{Task Assignment}, where decomposed subtasks are allocated to corresponding existing system monitors. In the \textit{Data Collection and Processing} phase, these monitors gather data to fulfill their assigned subtasks. Finally, during \textit{Task Merging}, the collected data is combined according to the original decomposition logic, effectively implementing the complete monitoring task while minimizing resource utilization.
This section elaborates the core mechanisms of this methodology, encompassing \textbf{Task Decomposition} and \textbf{Deployment}.

\subsubsection{Task Decomposition}
This phase establishes standardized constraints for decomposable tasks and introduces an algorithm for systematically decomposing complex monitoring tasks into manageable subtasks with integration logic. This integration logic enables subtasks to combine their monitoring outputs, reconstructing results for the original complex task.

Specifically, given a monitoring task $T$, we represent it symbolically as a triple $\langle V_T, O_T, T_T \rangle$, where:
\begin{itemize}
\item $V_T$ represents target system entities requiring observation by this monitoring task
\item $O_T$ represents property descriptions of target entities and their output types, $O_T = (a_T, t_T)$
\item $T_T$ represents system events requiring observation by this monitoring task
\end{itemize}

Based on this definition, the monitoring task decomposition problem can be formalized as: given a monitoring task $T$ and existing monitoring capabilities set $C = \{c_1, c_2, ..., c_n\}$ in the system, our objective is to identify a monitoring capabilities subset $C' \subseteq C$ and corresponding integration operations set $\lambda$, such that the combination satisfies task $T$ requirements.


\begin{algorithm}[tb]
\small
\caption{Task Decomposition DecomposeTask}
\label{alg:decomposition}
\begin{algorithmic}[1]
\Require Complex monitoring task $T$, Existing collector capabilities $C$
\Ensure Subtask set $S$, New collector requirements $N$, Integration logic $P$
\Function{DecomposeTask}{$T, C$}
\State $S \gets \emptyset$, $N \gets \emptyset$, $P \gets \emptyset$
\State $(T_{sub}, P_{compose}) \gets$ GenerateSubtasks($T$, $C$) \Comment{Subtasks and integration logic}
\For{each $t_i$ in $T_{sub}$}
\If{$\mu(t_i, C) = 1$}  \Comment{Mappable to existing collectors}
\State $S \gets S \cup \{t_i\}$
\State $P \gets P \cup$ GenerateIntegrationLogic($t_i, P_{compose}, C$)
\Else
\State $(S', N', P') \gets$ DecomposeTask($t_i, C$)
\If{$S' \neq \emptyset$} \Comment{If subtask set non-empty}
\State $S \gets S \cup S'$
\State $N \gets N \cup N'$
\State $P \gets P \cup P'$
\Else \Comment{Otherwise requires new collector}
\State $N \gets N \cup \{t_i\}$  \Comment{Requires new collector}
\State $P \gets P \cup$ GenerateIntegrationLogic($t_i$,$P_{compose}$)
\EndIf
\EndIf
\EndFor
\State \Return $(S, N, P)$
\EndFunction
\end{algorithmic}
\end{algorithm}

Based on these constraints, we propose Algorithm~\ref{alg:decomposition}, which processes complex monitoring task $T$ through recursive decomposition and integration. Initially, \emph{GenerateSubtasks($T$, $C$)} (line 3) decomposes $T$ into subtasks set $T_{sub}$ and integration logic $P_{compose}$.

Specifically, as Algorithm~\ref{alg:generatesubtasks} \textit{GenerateSubtasks} demonstrates, this algorithm describes an adaptive monitoring task decomposition and integration process. Algorithm~\ref{alg:generatesubtasks} first traverses existing collector set $C$ in lines 3-8, examining whether a single collector can directly satisfy monitoring task $T$ requirements, including target entity ($V_T$), attribute output type ($O_T$), and event type ($T_T$) coverage. When no directly satisfying collector is found, Algorithm~\ref{alg:generatesubtasks} introduces a large language model for task decomposition in lines 9-23, constructing prompts containing existing collector capability descriptions to generate integration logic $P_I'$ and corresponding monitoring capabilities $C_I'$. The algorithm verifies generated integration scheme feasibility; if satisfying original monitoring requirements, relevant subtasks are added to $T_{sub}$ and integration logic $P_{compose}$ is updated, otherwise up to three optimization iterations are performed. Finally, the algorithm outputs optimized subtask set $T_{sub}$ and corresponding integration logic $P_{compose}$, achieving automated monitoring task decomposition and dynamic integration.

\begin{algorithm}[tb]
\small
\caption{Generate Subtasks}
\begin{algorithmic}[1]
\Require Monitoring task $T$, Existing collector set $C$
\Ensure Subtask set $T_{sub}$, Integration logic $P_{compose}$
\Function{GenerateSubtasks}{$T, C$}
\State $T_{sub} \gets \emptyset$, $P_{compose} \gets \emptyset$
\For{each collector $c_j \in C$}
    \If{$T.V_T \subseteq c_j.V_c$ $\land$ $T.O_T \subseteq c_j.O_c$ $\land$ $T.T_T \subseteq c_j.T_c$} 
    
    \Comment{\textbf{\emph{Target entities covered by existing monitors}}}
            \State $T_{sub} \gets T_{sub} \cup \{T\}$
            \State $P_{compose} \gets P_{compose} \cup T_{sub}$
    \EndIf
\EndFor

\If{$T_{sub}$ is empty} \Comment{If no directly satisfying collector}
    \State $prompts \gets$ Create LLM Prompt 
    \For{each $c_j \in C$}
        \State $prompts \gets prompts + \{c_j\}$ 
        
        \Comment{\textbf{\emph{Add existing collector capability descriptions}}}
    \EndFor
     \State $prompts \gets prompts$ + "Task requirements: target entities $V_T$, attribute output type $O_T$, event type $T_T$, please generate integration logic that can combine existing monitoring capabilities to satisfy these requirements."

             \Comment{\textbf{\textit{Utilize LLM for task decomposition}}}
    \State $P_I',C_I' \gets$ LLM-generated integration logic, integrated monitoring capabilities 
    
    \If{$C_I'$ satisfies $T$'s target entities $V_T$, attribute output type $O_T$ and system events $T_T$}

        \Comment{\textbf{\textit{Determine if generated integration logic satisfies monitoring requirements}}}
        \State Add subtasks involved in $C_I'$ to $T_{sub}$ and generate integration logic
        \State $P_{compose} \gets P_{compose} \cup P_I'$
    \Else
        \State Request further optimization of integration logic (attempt 3 times)
    \EndIf
\EndIf

\State \Return $T_{sub}, P_{compose}$
\EndFunction
\end{algorithmic}
\label{alg:generatesubtasks}
\end{algorithm}

Subsequently, Algorithm~\ref{alg:decomposition} lines 4-19 evaluates each subtask using matching function $\mu(t_i, C)$, determining whether a monitoring task can be fulfilled by existing collectors:


\begin{equation}
\small
\mu(t_i, C) = \begin{cases}
1 & \text{if } \exists c_j \in C: t_i.V_T \subseteq c_j.V_c \\ 
  & \land\; t_i.O_T \subseteq c_j.O_c \land t_i.T_T \subseteq c_j.T_c \\
0 & \text{Otherwise}
\end{cases}
\end{equation}

Here, mappable subtasks ($\mu(t_i,C) = 1$, Algorithm~\ref{alg:decomposition} lines 5-7) are added to $S$ with corresponding integration logic via \emph{GenerateIntegrationLogic} (Algorithm~\ref{alg:decomposition} line 7), which derives integration operations $\lambda$ required for combining their outputs. The obtained integration logic is recorded in $P$.
Specifically, as Algorithm~\ref{alg:generate_integration_logic} lines 3-8 illustrate, the algorithm traverses existing basic information collectors $c_i \in C$, identifying collectors matching subtask $t_i$, and integrates them into existing integration logic $P_{compose}$ to form final integration logic $P$.

For unmappable subtasks (Algorithm~\ref{alg:decomposition} lines 8-18), we attempt further decomposition using the previously described method. If further decomposition is possible (line 10), resulting subtasks are recursively processed, with integration logic incorporated into $P$ (Algorithm~\ref{alg:decomposition} lines 11-13).

If further decomposition is impossible, the task is marked as requiring new collector capabilities (Algorithm~\ref{alg:decomposition} line 15, equivalent to Algorithm~\ref{alg:generate_integration_logic} lines 9-12), and its integration logic is added to $P$ (line 16). This process continues until all subtasks are either mapped to existing collectors or identified as new capability requirements in $N$.

\begin{algorithm}[tb]
\small
\caption{Generate Integration Logic }
\label{alg:generate_integration_logic}
\begin{algorithmic}[1]
\Require Subtask $t_i$, Existing integration logic $P_{compose}$, Optional parameter existing collector capabilities $C$ (default: None)
\Ensure Integration logic $P$
\Function{Generate\_Integration\_Logic}{$t_i, P_{compose}, C = \text{None}$}
\State $P \gets \emptyset$
\If{$C \neq \text{None}$} 
    \For{each basic information collector $c_i \in C$} 
        \If{collector $c_i$ corresponding to subtask $t_i$ found} 
            Update $P_{compose}$, appending corresponding collector $c_i$ to $t_i$ forming $P$
        \EndIf
    \EndFor
\Else \Comment{No existing basic information collectors available, must construct new ones}
    \State $P_{new} \gets$ CreateNewCollector($t_i$)  \Comment{Manually construct new monitor}
    \State Update $P_{compose}$, appending newly constructed $P_{new}$ to $t_i$ forming $P$
    \State \Return $P$
\EndIf
\EndFunction
\end{algorithmic}
\end{algorithm}


\noindent\textbf{Cost Rank}
To formalize our approach to cost optimization within the security monitoring framework, we introduce a comprehensive cost function that quantifies the tradeoffs involved in monitoring task decomposition. This function serves as a critical component in our optimization process, guiding the selection of implementation strategies that minimize resource utilization while maintaining effective security coverage.

The cost function $C(T)$ for a monitoring task $T$ decomposed into subtasks $\{t_1, t_2, ..., t_n\}$ is formulated as:

\begin{equation}
\small
C(T) = \sum_{i=1}^{n} \big(D_{\text{deploy}}(t_i) + D_{\text{dev}}(t_i)\big) + C_{\text{complex}}(T)
\end{equation}

Where $D_{\text{deploy}}$ denotes deployment costs, $D_{\text{dev}}$ denotes development costs, and $C_{\text{complex}}$ captures integration complexity. 

For deployment costs, we differentiate between existing and new monitoring components:

\begin{equation}
\small
D_{\text{deploy}}(t_i) = 
\begin{cases}
\alpha \cdot P_{\text{overhead}}(t_i) & \text{if } t_i \text{ maps to existing monitor} \\
\beta_{\text{imp}} \cdot P_{\text{overhead}}(t_i, \text{imp}) & \text{if new monitor required}
\end{cases}
\end{equation}

Where $\alpha$ is a weighting factor for existing monitors, $P_{\text{overhead}}$ quantifies performance impact, and $\beta_{\text{imp}}$ represents implementation-specific weights that vary according to implementation approach (hardware, kernel, or user-space). Notably, the relationship $\beta_{\text{hw}} < \beta_{\text{kernel}} < \beta_{\text{user}}$ reflects our observation that hardware implementations typically introduce less runtime overhead than kernel-level implementations, which in turn impact performance less than user-space implementations.


Development costs are structured to minimize resource usage by prioritizing existing capabilities: zero cost for reusing existing monitors, with increasing costs for user-space, kernel-level, and hardware implementations respectively (i.e., $0 < \gamma_{\text{user}} < \gamma_{\text{kernel}} < \gamma_{\text{hw}}$).

With weighting factors satisfying $\gamma_{\text{hw}} > \gamma_{\text{kernel}} > \gamma_{\text{user}}$, reflecting the relative development effort associated with each implementation approach. This formulation encourages the reuse of existing monitoring capabilities when possible, as these components incur zero additional development cost.

Integration complexity is modeled using a logarithmic function to reflect the sub-linear growth in complexity as the number of components increases, i.e., $C_{\text{complex}}(T) = n$, 
Where $n$ represents the number of subtasks in the decomposition.


After generating potential implementation solutions through task decomposition, we evaluate each solution using the cost function defined above. Let $S = \{S_1, S_2, ..., S_m\}$ represent the set of candidate solutions, where each solution $S_j$ consists of a specific decomposition of the original monitoring task. The cost evaluation function assigns a numerical cost score to each solution, i.e., $\text{Score}(S_j) = C(S_j)$


We then rank all candidate solutions by their cost scores in ascending order.
The solution with the minimum cost score is selected as the optimal implementation strategy. 
This process ensures that our monitoring implementation balances comprehensive security coverage with practical resource constraints. By systematically evaluating and comparing different implementation strategies, we achieve optimal resource utilization while maintaining effective security monitoring capabilities.

\subsubsection{Deployment}
After systematically decomposing monitoring tasks and optimizing their implementation strategy through our cost-aware approach, we proceed to the deployment phase. This phase encompasses three key stages: (1) \textbf{Task Assignment} to appropriate collectors, (2) \textbf{Data Collection and Processing} according to task specifications, and (3) integration of collected data through \textbf{Task Merging}. 

\smallskip
\noindent\textbf{Task Assignment.} Based on $\mu(t_i, C)$, subtasks $t_i \in S$ are assigned to existing collectors $C$ or marked as new collector requirements $N$. Subtasks with $\mu(t_i, C) = 1$ are mapped to compatible collectors for direct information collection, while subtasks with $\mu(t_i, C) = 0$ are added to $N$, and corresponding custom collectors are implemented through expert intervention to fulfill monitoring requirements.

\smallskip
\noindent\textbf{Data Collection and Processing.} Collectors execute tasks based on $t_i$, producing data outputs ${O_t}$.

\smallskip
\noindent\textbf{Task Merging.} Using integration logic $P$ from \emph{Task Decomposition}, collected outputs $\{O_t\}$ are combined to reconstruct original task result $O_T$. Integration operations $\lambda \in P$ ensure consistency with $T$'s output requirements.

Through this systematic approach, our methodology maximizes utilization of existing collectors, reducing requirements for new modules and deployments, thereby optimizing resource efficiency.

\subsection{Monitor Construction and Deployment Case Study}
\label{sec:monitorcasestudy}



\begin{figure}
\centering
\includegraphics[width=0.5\textwidth]{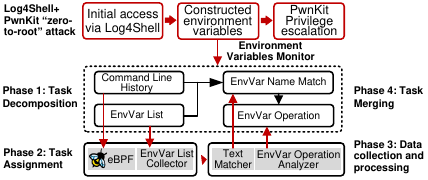}
\caption{SMART method Log4Shell Zero-to-Root Attack example.}
\label{fig:task}
\end{figure}

Consider a \emph{Log4Shell zero-to-root} attack scenario (Figure~\ref{fig:task}), where attackers exploit the Log4Shell vulnerability for initial access and then manipulate environment variables (EnvVar) for privilege escalation. A critical monitoring requirement is tracking \emph{environment variable modifications}. Traditional security monitoring relies on system call events, but this approach has major limitations: environment variable operations typically execute through shell built-in functions that don't trigger system calls. This creates monitoring blind spots since traditional system call tracking cannot detect these operations. As environment variable manipulation is often critical in privilege escalation attacks, these blind spots significantly impact system security awareness capabilities.

Therefore, based on the real-world attack scenario analysis, we must enhance the existing security monitoring framework to address these critical requirements. Using Algorithm~\ref{alg:decomposition}'s task decomposition methodology, we decompose \textit{environment variable modification monitoring} into fundamental implementable subtasks. Specifically, we decompose it into two basic subtasks $T_{sub}$: command history monitoring ($t_1$) that tracks command line activities with output type $O_{t1} = \{e_1, ..., e_n\}$, where each $e_i$ contains command string, process ID and timestamp; and environment variable list monitoring ($t_2$) that tracks the system's environment variable names, with output type $O_{t2} = \{v_1, ..., v_m\}$, where each $v_i$ represents an environment variable name.


This task decomposition implements two key monitoring components: $t_1$ is responsible for command line history monitoring, implemented through eBPF technology. Specifically, we utilize eBPF's dynamic tracing capabilities to perform probe instrumentation at shell program key functions (such as readline, execute\_command, etc.), capturing user input command sequences in real-time. $t_2$ focuses on environment variable monitoring, where we developed an independent collector program that periodically obtains and records system environment variable snapshots at predetermined intervals.

The integration logic $P_{compose}$ implements data association analysis as follows: first through operation $\lambda_{name}(O_{t1}, O_{t2}) = \{cmd \in O_{t1} | \exists var \in O_{t2}: var \text{ appears in } cmd.cmd\_str\}$, it performs name matching between command strings and environment variable names, then analyzes matched commands to identify current environment variable operations.

\subsection{The Anomaly Detection Module}

Building upon our observations of attack locality, we introduce a weighted mechanism with attack locality awareness to enhance threat detection effectiveness. This mechanism leverages endpoint anomaly information embedded in the provenance graph mentioned earlier and guides weight allocation through anomaly scoring, thereby improving the identification and response to potential threats. The following sections detail our approach through three stages: provenance graph node embedding, graph neural network design, and data post-processing.

\smallskip
\noindent\textbf{Provenance Graph Node Embedding.} Given a provenance graph $G = (V, E)$ where $V = \{v | v \in (Process \cup File \cup Socket)\}$ and $E \subseteq V \times V \times T$, each vertex $v \in V$ is characterized by a feature vector that encodes its entity-behavior interaction patterns:

\begin{equation}
\small
h_v = [f_{in}(v) \parallel f_{out}(v) \parallel S(v)]
\end{equation}

where $f_{in}(v)$ and $f_{out}(v)$ represent the distributions of incoming and outgoing edge types (i.e., behavior types) respectively, and $S(v)$ is the anomaly score of node $v$ as described in Section \ref{sec:pre}.

\begin{equation}
\small
\label{equ:fin}
f_{in}(v)[t] = |{(u,v,type) \in E | type=t, t \in T}|
\end{equation}

\begin{equation}
\small
\label{equ:fout}
f_{out}(v)[t] = |{(v,u,type) \in E | type=t, t \in T}|
\end{equation}

Specifically, $f_{in}(v)[t]$ (Equation \ref{equ:fin}) counts the number of edges of type $t$ that are directed toward node $v$. 
This captures the frequency of specific types of behaviors performed by other nodes on $v$.
Similarly, $f_{out}(v)[t]$ (Equation \ref{equ:fout})
counts how many edges of type $t$ originate from node $v$. This captures the frequency of specific types of behaviors that node $v$ performs on other nodes.

In essence, these formulas compute the frequency of each behavior type associated with node $v$, categorized as incoming (behaviors from other nodes toward $v$) and outgoing (behaviors from $v$ toward other nodes). The resulting feature vector encodes the behavioral patterns of nodes in the provenance graph, aiding in the identification of anomalous or potentially malicious activities.

Furthermore, the inclusion of the anomaly score $S(v)$ enhances the discriminative capability of the feature representation, effectively guiding branch weight allocation during the global provenance analysis process, thereby improving the overall accuracy and reliability of attack detection.

\smallskip
\noindent\textbf{Graph Neural Network Design.} In this paper, we use a two-layer Graph Attention Network (GAT)~\cite{velivckovic2018graph} to learn vertex features while capturing vertex-edge (i.e., entity-behavior) patterns. During the learning process, the feature update for vertex $i$ is computed by:

\begin{equation}
\small
h_i = \sum_{j \in \mathcal{N}(i)} \alpha_{ij} \mathbf{W}_2 \text{ELU}\left(\sum_{k \in \mathcal{N}(j)} \alpha_{jk} \mathbf{W}_1 h_k \right)
\end{equation}

Here, $\mathbf{W}_1$ and $\mathbf{W}_2$ are the weight matrices for the first and second layers. The attention coefficients $\alpha_{ij}$ ($\alpha_{jk}$) determine the importance of neighbor $j$'s ($k$'s) features to vertex $i$ ($j$). Exponential Linear Unit (ELU) serves as the inter-layer activation function, which enhances gradient flow, improves expressive capability, and provides non-linear transformations across negative value ranges, thereby improving model performance.

For anomaly detection, we first train our GAT on benign provenance graphs. By adjusting the vertex weights based on vertex-edge (i.e., entity-behavior) patterns in benign graphs, we obtain a benign GAT entity-behavior model. In the prediction phase, we apply a multi-class classification approach to the output of the GAT layer. Specifically, after feature aggregation, we apply the \emph{softmax} function to the vertex feature $h_v$ to produce class probabilities, i.e., $z_v = \text{softmax}(h_v)$. Subsequently, we employ the argmax function to obtain prediction results from the softmax output. Specifically, we select the class with the highest probability as the model's final predicted class, i.e., $\hat{y}_v = \arg\max_{c} z_v^c$, where $z_v^c$ represents the probability that vertex $v$ belongs to class $c$.

\smallskip
\noindent\textbf{Data Post-processing.} We define $is\_anomalous(v)$ to check vertex $v$. If its predicted entity type differs from its actual type, we consider it anomalous, having deviated from our trained benign entity-behavior model. When anomalies are detected, we analyze their neighbors, where attention weights accumulate among connected anomalous vertices while becoming diluted among benign ones.

To further reduce false positives, for each predicted anomalous vertex $v$, we analyze its $k$-hop neighborhood $N_k(v)$ (where $k=2$ in our implementation) and compute a benign density score:

\begin{equation}
\small
\text{Benign Density}(v) = \frac{| \{ u \in N_k(v) : \neg \text{is\_anomalous}(u) \} |}{|N_k(v)|}
\end{equation}

If $\text{Benign Density}(v)$ exceeds a threshold (80\% in our implementation), indicating that the majority of its neighboring vertices are benign, we consider $v$ to lack attack locality and correct it to benign, further reducing false positives.

\section{Evaluation.}
\label{sec:eva}
We evaluate \tool by answering the following research questions:

\noindent $\bullet$ \textbf{RQ1: (Monitoring Coverage)} What security-critical monitoring items from real-world attacks are captured, and do existing tools miss any of these?

\noindent $\bullet$ \textbf{RQ2: (Effectiveness)} How effective is \tool in detecting attack events and anomalies in resource-constrained environments?

\noindent $\bullet$ \textbf{RQ3: (Ablation Study)} How do the components and design choices of \tool impact its effectiveness in attack detection and anomaly identification?

\noindent $\bullet$ \textbf{RQ4: (Deployment Costs)} What are \tool's development costs, and is it feasible for real-world deployment?





\smallskip
\begin{figure*}[t]
\centering
\includegraphics[width=0.9\textwidth]{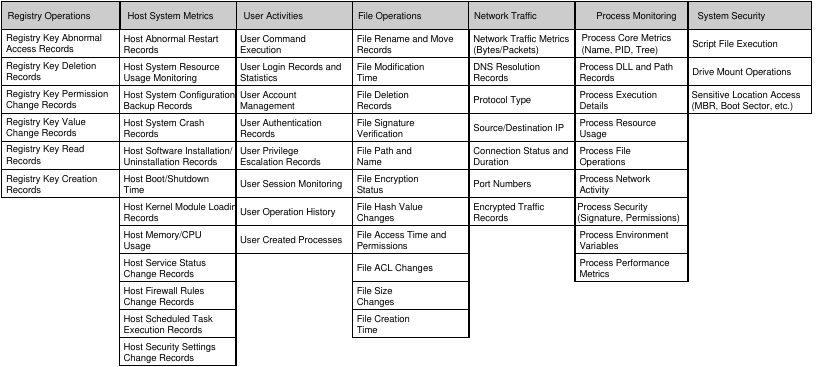}
\caption{Security-focused monitoring items from real-world attacks.}
\label{fig:monitor}
\end{figure*}

\noindent\textbf{Evaluation Datasets}
We evaluate our approach using the DARPA TC~\cite{darpa} and CSE-CIC-IDS2018~\cite{cse} datasets, as in previous work~\cite{wang2022threatrace, zengy2022shadewatcher, alsaheel2021atlas, lo2022graphsage}. The DARPA TC dataset consists of four scenarios—THEIA, Trace, CADETS, and Fivedirections—covering various attack steps and environments. The CSE-CIC-IDS2018 dataset, from the Canadian Institute for Cybersecurity (CIC), includes data on various attacks, such as Brute Force, Heartbleed, etc. Table \ref{tab:existing_datasets} summarizes these datasets.

\begin{table}[!tbp]
\centering
\caption{Evaluation Datasets Overview}
\scalebox{0.75}{
\begin{tabular}{|c|c|c|c|c|}
\hline
\textbf{Dataset} & \textbf{Scenario} & \textbf{Benign Vertices} & \textbf{Anomalous Vertices} & \textbf{Edges} \\
\hline
\multirow{4}{*}{DARPA TC} & THEIA & 3,505,326 & 25,362 & 102,929,710 \\
& Trace & 2,416,007 & 67,383 & 6,978,024 \\
& CADETS & 706,966 & 12,852 & 8,663,569 \\
& Fivedirections & 569,848 & 425 & 9,852,465 \\
\hline
CSE-CIC-IDS2018 & - & 212,628 & 89,228 & 501,856 \\
\hline
\end{tabular}
}
\label{tab:existing_datasets}
\end{table}

Since existing datasets do not incorporate our constructed monitoring system, we create custom datasets to evaluate \tool’s monitoring designs and anomaly detection capabilities. We collaborate with two industry experts to reproduce common exploits, such as the Log4j vulnerability (CVE-2021-44228)\cite{wu2023effective, chen2024exploiting, haq2024sok} and the OpenSMTPD vulnerability (CVE-2020-7247)\cite{ruan2024vulzoo, wang2023research, becker2024evaluation}, along with the associated attack activities based on real-world application scenarios.
The datasets include: \textbf{(1) Log4j+ENV Attack Dataset:}
This dataset simulates attacks related to the Log4j vulnerability, covering scenarios like initial access to the target, privilege escalation via environment variables, and the establishment of reverse shell connections.
\textbf{(2) OpenSMTPD+Malicious Execution:}
This dataset demonstrates an attack chain exploiting the OpenSMTPD vulnerability, including unauthorized command execution, downloading malicious scripts, and executing additional malware.

\smallskip
\noindent\textbf{Deployment Environment.}
\textbf{(1) For attack report analysis and security monitoring item extraction}, we use Microsoft’s Azure OpenAI API~\cite{azure} with the GPT-3.5-16K model
and set the temperature parameter to 0.4 to balance creativity and accuracy.
\textbf{(2) For our GAT implementation}, we use a two-layer GAT architecture with 8 attention heads and a hidden layer of 128 units. We set the batch size to 500, learning rate to 0.01, weight decay to 5e-4, and a dropout rate of 0.5 to prevent overfitting.
\textbf{(3) For \tool deployment and anomaly detection experiments}, we use Debian 10.8 (Linux 5.10) with an Intel(R) Core(TM) i5-12500 processor and 48GB RAM, hosting all information collection tools and attack detection systems.
\textbf{(4) For our cost function implementation}, we determined parameters through expert evaluation. Four co-authors with 2-3 years of attack detection research experience, plus two industry experts with 7-8 years of cybersecurity experience, collaboratively established parameter values using majority voting to resolve disagreements. We set $\alpha = 0.2$ for existing monitor weights, reflecting minimal overhead when reusing components. For new implementations, we assigned $\beta_{user} = 0.7$, $\beta_{kernel} = 0.5$, and $\beta_{hw} = 0.3$, capturing our observation that user-space implementations typically introduce higher runtime overhead than kernel-space, while hardware-accelerated monitors showed the lowest impact. Development costs were parameterized as $\gamma_{user} = 10$, $\gamma_{kernel} = 25$, and $\gamma_{hw} = 50$, representing relative implementation effort in person-days based on previous projects, with hardware implementations requiring more specialized expertise and time. This methodology ensured our cost function balanced theoretical soundness with practical considerations, producing task decompositions that maximized existing monitor utilization while minimizing development overhead.

\begin{table}[!tbp]
\small
\centering
\setlength{\tabcolsep}{4.5pt} 
\renewcommand{\arraystretch}{1} 
\caption{Custom Dataset Information}
\scalebox{0.75}{
\begin{tabular}{|c|c|c|c|}
\hline
\textbf{Scenario} & \textbf{Benign Vertices} & \textbf{Anomalous Vertices} & \textbf{Edges} \\
\hline
Log4j+ENV & 812 & 37 & 1,975 \\
\hline
OpenSMTPD+Malicious Execution & 3,384 & 34 & 7,989 \\
\hline
\end{tabular}}
\label{tab:custom_datasets}
\end{table}


\begin{table*}[t]
\setlength{\tabcolsep}{13.7pt} 
\caption{Comparison of detection effectiveness}
\label{tb:comparison}
\centering
\begin{tabular}{|l|c|c|c|c|c|c|c|c|}
\hline
\multirow{2}{*}{Dataset} & \multicolumn{4}{c|}{Our Approach} & \multicolumn{4}{c|}{ThreaTrace} \\ 
\cline{2-9}
& Precision & Recall & FPR & F1-Score & Precision & Recall & FPR & F1-Score \\ 
\hline
Cadets         & 97.92\% & 99.88\% & 0.08\%  & \textbf{98.89\%} & 93.84\% & 99.96\% & 0.24\%  & 96.81\% \\ 
Fivedirections & 72.53\% & 95.06\% & 0.04\%  & \textbf{82.28\%} & 75.21\% & 84.94\% & 0.032\% & 79.78\% \\ 
Theia          & 99.81\% & 99.91\% & 0.02\%  & \textbf{99.86\%} & 95.49\% & 99.90\% & 0.37\%  & 97.65\% \\ 
Trace          & 98.24\% & 99.996\% & 0.02\% & \textbf{99.11\%} & 81.59\% & 99.99\% & 1.33\%  & 89.86\% \\ 
\hline
CSE-CIC        & 98.99\% & 93.55\%  & 0.04\% & \textbf{96.19\%} & 92.58\% & 95.12\% & 7.58\%  & 93.44\% \\ 
\hline
Log4j+ENV      & 99.46\% & 100\%    & 0.02\%      & \textbf{99.73\%} & 76.09\% & 99.99\% & 1.36\%  & 86.42\% \\ 
OpenSMTPD      & 94.97\% & 100\%    & 0.05\%      & \textbf{97.42\%} & 77.50\% & 91.18\% & 0.27\%  & 83.78\% \\ 

\hline
Average        & 94.41\%   & 98.91\%  & 0.03\% & \textbf{96.76\%} & 84.61\%     & 95.87\% & 1.57\%  & 88.53\% \\ 
\hline
\end{tabular}
\end{table*}

\subsection{RQ1: Monitor Items and Monitoring Coverage Evaluation}

Fig. \ref{fig:monitor} presents the 85 security-relevant monitoring items our methodology has extracted from real-world attack reports and ATT\&CK cases. While traditional monitoring tools primarily focus on basic system elements (process PIDs, names, arguments, file paths/names, and network IPs and ports).
In contrast, our approach systematically broadens the monitoring scope. Beyond these fundamental elements, our framework incorporates detailed monitoring of system security configurations (e.g., firewall rule changes, security setting modifications, and service status updates), user authentication patterns (e.g., login records, privilege escalation events, and session tracking), process behaviors (e.g., DLL loading, resource usage, and network activity), and file integrity metrics (e.g., hash values, encryption status, and changes to access control lists), which are frequently neglected by conventional monitoring tools.

As shown in Table~\ref{tab:coverage}, while \textit{Auditd} and \textit{Camflow} are widely used to generate system logs for provenance graphs, they were not originally designed for security, resulting in low coverage—49.40\% and 40.00\% respectively—for the monitoring requirements derived from real-world attacks. \textit{Auditbeat}, Elastic's security-enhanced version of \textit{Auditd}, demonstrates how expert knowledge can close these gaps, achieving 75.30\% coverage.
More recently, eBPF has gained traction for its low overhead and real-time capabilities in Linux environments. We evaluated \textit{eAudit}, an academic extension of \textit{Auditd} using eBPF, which improved coverage to 56.47\%, though with room for further improvement.
Focusing on Linux (excluding Windows-specific tools like ETW), our approach applies the methodology from Section~\ref{sec:secmonitor} (with case studies in Section~\ref{sec:monitorcasestudy}). By decomposing complex monitoring tasks, leveraging existing monitors, and integrating collected data, we extended coverage to 83.50\%. We further validated the effectiveness and practicality of our monitoring framework through ablation experiments in \textit{\textbf{RQ3}}.

\begin{table}[tb!]
\small
\setlength{\tabcolsep}{2pt}
\renewcommand{\arraystretch}{1}
\caption{Coverage Analysis of Different Monitoring Tools}
\label{tab:coverage}
\centering
\scalebox{0.89}{
\begin{tabular}{|c|c|c|c|c|c|c|}
\hline
\textbf{Metrics} & \textbf{Auditd} & \textbf{Auditbeat} & \textbf{eAudit} & \textbf{Camflow} & \textbf{Our Method} \\
\hline
Coverage Quantity & 42 & 64 & 48 & 34 & 71 \\
\hline
Coverage Rate (\%) & 49.40\% & 75.30\% & 56.47\%
& 40.00\% & 83.50\% \\
\hline
\end{tabular}
}
\end{table}

\subsection{RQ2: Effectiveness of \tool}

\begin{table*}[tb!]
\caption{Ablation Study: Monitoring Framework and Attack Locality}
\setlength{\tabcolsep}{13.3pt} 
\label{tab:ablation}
\centering
\begin{tabular}{|c||c|c|c|c||c|c|c|c|}
\hline
\multirow{2}{*}{Data Source} & \multicolumn{4}{c||}{With Our Monitoring Framework} & \multicolumn{4}{c|}{Without Our Monitoring Framework} \\
\cline{2-9}
& Precision & Recall & FPR & F1-Score & Precision & Recall & FPR & F1-Score \\
\hline
opensmtpd & 94.97\% & 100\% & 0.05\% & \textbf{97.42\%} & 78.57\% & 97.06\% & 0.27\% & 86.84\% \\
\hline
log4jEnv & 99.46\% & 100\% & 0.02\% &\textbf{99.73\%} & 74.00\% & 100\% & 1.6\% & 85.06\% \\
\hline
Average & 97.22\% & 100\% & 0.035\% & \textbf{98.58\%} & 76.29\% & 98.53\% & 0.94\% & 85.95\% \\
\hline
\hline
\multirow{2}{*}{Scenario} & \multicolumn{4}{c||}{With Attack Locality} & \multicolumn{4}{c|}{Without Attack Locality} \\
\cline{2-9}
& Precision & Recall & FPR & F1-Score & Precision & Recall & FPR & F1-Score \\
\hline
Cadets & 97.92\% & 99.88\% & 0.08\% & \textbf{98.89\%} & 84.14\% & 99.89\% & 0.70\% & 91.34\% \\
\hline
Fivedirections & 72.53\% & 95.06\% & 0.04\% & \textbf{82.28\%} & 44.30\% & 95.06\% & 0.14\% & 60.43\% \\
\hline
Theia & 99.81\% & 99.91\% & 0.02\% & \textbf{99.86\%} & 92.11\% & 99.91\% & 0.68\% & 95.85\% \\
\hline
Trace & 98.24\% & 99.996\% & 0.02\% & \textbf{99.11\%} & 89.47\% & 99.995\% & 0.11\% & 94.44\% \\
\hline
Average & 92.13\% & 98.71\% & 0.04\% & \textbf{95.04\%} & 77.51\% & 98.71\% & 0.41\% & 85.52\% \\
\hline

\end{tabular}
\end{table*}

To evaluate the effectiveness of our proposed approach, we conducted comprehensive experiments on multiple datasets and compared our method against ThreaTrace~\cite{wang2022threatrace}, a state-of-the-art GraphSAGE-based detection method for provenance graph anomaly detection that has received significant citations and provides complete open-source implementation.

Table~\ref{tb:comparison} shows that our approach demonstrates superior detection performance (average 8.23\% higher F1-score) compared to ThreaTrace. A detailed analysis of performance across individual datasets reveals the following improvements:


\noindent\textbf{DARPA Dataset Performance Analysis:} Our method demonstrates remarkable consistency across the four \textit{DARPA datasets}, achieving notable improvements on the \textit{Theia} dataset with up to 9.25\% higher F1-scores and up to 1.31\% lower false positive rates (FPR). This translates to a 16.65\% precision increase, highlighting our locality-aware anomaly detection approach's effectiveness in established complex system environments.
Similarly impressive results on the \textit{Trace} dataset show that our method achieved an outstanding 99.11\% F1-score compared to ThreaTrace's 89.86\%, primarily due to our substantially higher precision (98.24\% versus 81.59\%) while maintaining comparable recall. These improvements demonstrate how our attack locality-based vertex weighting mechanism effectively distinguishes between benign and malicious activities in diverse system behaviors.

\noindent\textbf{CSE-CIC Dataset Performance Analysis:} On this dataset, our method achieves a 96.19\% F1-score, outperforming ThreaTrace's 93.44\%. Most remarkably, our approach drastically reduces the FPR to merely 0.04\% compared to ThreaTrace's 7.58\% - a 189-fold improvement that would significantly reduce the number of false alarms security analysts must investigate. This substantial performance gap further validates that our locality-aware anomaly detection mechanism effectively leverages the clustering characteristics of malicious activities, providing more precise differentiation between normal network traffic and genuine attacks.

\noindent\textbf{Custom Attack Dataset Performance Analysis:} To validate our security monitoring framework's effectiveness, we collected data from real-world attacks using our proposed monitoring system. Here, our approach shows remarkable improvements, with the Log4j+ENV dataset showing a 13.31\% higher F1-score compared to ThreaTrace. Our approach achieves a near-perfect 99.73\% F1-score versus ThreaTrace's 86.42\%, demonstrating exceptional detection capability for this sophisticated attack vector. For the OpenSMTPD dataset, we observe similarly impressive results, reaching 97.42\% F1-score while maintaining a remarkably low 0.05\% FPR. 
These results on modern attack scenarios clearly show how our combined approach - using both security-oriented monitoring framework and locality-aware analysis - successfully captures important attack signatures while correctly distinguishing benign nodes from the provenance graph.

\smallskip
These results confirm that our approach effectively combines enhanced security monitoring with locality-aware anomaly detection to significantly improve detection accuracy while reducing false positives across diverse environments.

\begin{table}[tb!]
\setlength{\tabcolsep}{1pt} 
\caption{Deployment Cost Analysis}
\centering
\label{tab:overhead}
\scalebox{0.78}{
\begin{tabular}{|l|l|l|l|}
\hline
\textbf{Metrics} & \textbf{Baseline} & \textbf{\tool (Cost)} & \textbf{Non-\tool (\textbf{Cost}}) \\
\hline
SPEC2006 (Execution time) & 2,724.6s & 2,871.85s (\textbf{\textcolor{red}{5.40\%}}) &  3043.65s (\textbf{\textcolor{red}{11.71\%}}) \\
STREAM (Throughput) & 19,036.75 MB/s & 18,988.28 MB/s (\textbf{\textcolor{red}{0.26\%}}) & 18860.30 MB/s (\textbf{\textcolor{red}{0.94\%}})
 \\
Application (Execution time) & 4,498ms & 4,514ms (\textbf{\textcolor{red}{0.36\%}}) &  5,814ms (\textbf{\textcolor{red}{29.26\%}})\\
\hline
Lines of Code & - & 59,203 & 86,257\\
\hline
\end{tabular}
}
\end{table}


\subsection{RQ3: Ablation Study of \tool}




To evaluate the effectiveness of our key design components, we conducted two ablation experiments: \textbf{(1) whether utilizing our security-oriented monitoring items framework enhances detection capabilities} - testing \tool's detection performance on our reproduced Log4j and OpenSMTPD vulnerability exploitation scenarios, with and without our monitoring items for data/log collection, and \textbf{(2) whether considering attack locality improves detection performance} - evaluating \tool's detection effectiveness on the widely-used DARPA datasets with and without attack locality consideration.

\smallskip
\noindent\textbf{Monitoring Framework Impact Analysis.} As shown in Table \ref{tab:ablation}, our monitoring items framework significantly improves detection capability, achieving an average F1-score of 98.58\% compared to 85.95\% without it, showing a 12.63\% improvement. The FPR drops by 0.91\% with our framework, corresponding to a 20.93\% precision increase. Examining the individual datasets reveals that the most dramatic improvement occurs on the Log4j+ENV scenario, where precision increases from 74.00\% to 99.46\% (a 25.46\% gain) while maintaining perfect recall. Similarly, for OpenSMTPD, precision increases from 78.57\% to 94.97\% (a 16.4\% improvement). These results validate that our systematically extracted monitoring items from real-world attacks are crucial for accurate attack detection, providing visibility into attack behaviors that would otherwise be missed.

\smallskip
\noindent\textbf{Attack Locality Impact Analysis.} \tool achieves an average F1-score of 95.04\% when considering locality patterns, versus 85.52\% without them. This demonstrates locality's effectiveness in improving detection precision by 9.52\% while maintaining a significantly lower false positive rate (0.04\% vs 0.41\%). The impact is most pronounced on the Fivedirections dataset, where F1-score increases from 60.43\% to 82.28\% (a 21.85\% improvement), primarily through enhanced precision (44.30\% to 72.53\%). Even on datasets where our approach already performs well, such as Theia (F1-score improvement from 95.85\% to 99.86\%), incorporating locality patterns further reduces false positives by 0.66\%. This consistent pattern across all datasets demonstrates that attack locality is a fundamental characteristic that effectively distinguishes genuine attacks from isolated anomalies.

These results confirm that both our security-oriented monitoring framework and attack locality consideration substantially enhance detection effectiveness while minimizing false alarms. This framework ensures comprehensive visibility into security-relevant system activities, while the locality-based analysis differentiates between benign anomalies and actual attack patterns, creating a multiplicative rather than merely additive improvement in overall detection capabilities.

\subsection{RQ4: Deployment Cost Analysis}



To evaluate the deployment costs of our approach, we measured both performance overhead and development complexity. For performance metrics, we assessed CPU performance using the widely-recognized SPEC2006 benchmark~\cite{spec,hassan2021reusable}, memory throughput using STREAM benchmark~\cite{stream,peng2020memory}, and application runtime overhead through instrumentation tests across 100 runs. For development complexity, we measured implementation effort in lines of code. While our FEAD approach decomposes complex monitoring tasks and distributes them to existing collectors, the non-FEAD implementation requires developing new monitoring capabilities from scratch.

Table \ref{tab:overhead} shows that \tool incurs minimal deployment costs. It introduces only a 5.40\% overhead in CPU performance (SPEC2006), a 6.31\% improvement over traditional methods. Memory throughput impact is even smaller, with a 0.26\% degradation, sustaining 18,988.28 MB/s. Most notably, \tool introduces a mere 0.36\% application runtime overhead, reduced by 28.90\%. In terms of development effort, \tool requires 59,203 lines of code, cutting implementation complexity by 31.4\%.
These results highlight \tool's monitoring strategy as an effective solution that minimizes both performance impact and development effort, making it highly suitable for production environments.
\section{Conclusion}
We proposed \tool, a novel framework for detecting sophisticated cyber threats in resource-constrained systems. We introduced an attack model-driven monitoring items identification approach that systematically extracts security-critical items from attack reports, proposed an efficient monitoring framework deployment via a complex task decomposition mechanism, and developed a locality-aware anomaly analysis technique that leverages the clustering characteristics of malicious activities. We conducted extensive evaluations on multiple real-world datasets and custom attack scenarios. The experimental results demonstrate that \tool outperforms existing solutions with an 8.23\% higher F1-score while maintaining only 5.4\% overhead, validating its effectiveness for efficient and accurate attack detection.

\section{ACKNOWLEDGMENT}
We are grateful for all the anonymous reviewers. The authors from Nanjing University are supported in part by the Leading-edge Technology Program of Jiangsu Natural Science Foundation (No. BK20202001), the National Natural Science Foundation of China (No. 62232008, 62172200), and the Postgraduate Research \& Practice Innovation Program of Jiangsu Province (No. KYCX24\_0237)

\clearpage
\section*{Ethical Considerations}
This research followed ethical guidelines throughout all phases of data collection and analysis. In Section \ref{sec:monitor}, threat intelligence was gathered exclusively from publicly accessible web pages while fully complying with site policies and ethical web crawling practices, ensuring no unauthorized access occurred. The evaluation in Section \ref{sec:eva} utilized publicly available DARPA and CSE-CIC datasets, which are widely accepted benchmarks in the security research community for assessing attack detection capabilities without introducing new ethical concerns. For the specialized dataset developed in Section \ref{sec:eva}, all domain experts were clearly informed about our research goals before participating, received strong assurances that no user privacy would be compromised, and explicit acknowledgment of their right to withdraw participation at any time without consequence, thus maintaining ethical integrity throughout the study.
\balance
\bibliographystyle{IEEEtran}
\bibliography{ref}

\end{document}